\documentclass[aps,onecolumn,superscriptaddress,amsmath,amssymb,showpacs]{revtex4}

\usepackage{graphicx}
\usepackage{dcolumn}
\usepackage{bm}
\usepackage{mathrsfs}
\usepackage{appendix}
\usepackage{bbm}
\usepackage{color}

\def \vU {{\bf T}}

\def \ve {{\bf e}}

\def \vm {{\bf m}}

\def \Cpe {C_{\perp}}
\def \Cpa {C_{\parallel}}
\def \Df {\textbf {\emph{D}}_{\rm FE}}

\def \Da {\textbf {\emph{D}}_{\rm AFD}}

\def \ni {{n_i}}

\def \vS {{\bf S}}
\def \vR {{\bf R}}
\def \ve {{\bf e}}
\def \vQ {{\bf Q}}

\def \vh {{\bf h}}
\def \vm {{\bf m}}

\def \mb {\mu_{\rm B}}

\def \psc {{\bf P}^{\rm SC}}

\def \cV {{\cal V}}

\def \xp {{\bf x}^{\prime }}
\def \yp {{\bf y}^{\prime }}
\def \zp {{\bf z}^{\prime }}

\def \xx {{\bf x}}
\def \yy {{\bf y}}
\def \zz {{\bf z}}

\def \vM {{\bf M}}
\def \vP {{\bf P}}
\def \vW {{\bf W}}
\def \BF {{\rm BiFeO$_3$} }
\def \BP {{\rm BiFeO$_3$}}

\def \TC {T_{\rm C}}
\def \TN {T_{\rm N}}
\def \vE {{\bf E}}

\def \xq {x^{\prime}}
\def \yq {y^{\prime}}
\def \zq {z^{\prime}}
\def \ds {\displaystyle }

\def \ve {{\bf e}}
\def \vh {{\bf h}}

\def \mB {\mu_{\rm B}}

\begin{document}

\title{First-principles approach to the dynamic magnetoelectric couplings in BiFeO$_3$\footnote{
This manuscript has been written by UT-Battelle, LLC under Contract No. DE-AC05-00OR22725 
with the U.S. Department of Energy.  The United States Government retains and the publisher, 
by accepting the article for publication, acknowledges that the United States Government retains 
a non-exclusive, paid-up, irrevocable, world-wide license to publish or reproduce the published form 
of this manuscript, or allow others to do so, for United States Government purposes.  The Department of Energy 
will provide public access to these results of federally sponsored research in accordance with the DOE Public Access Plan.
}}

\author{Jun Hee Lee*}
\affiliation{Materials Science and Technology Division, Oak Ridge National Laboratory, Oak Ridge, Tennessee, 37831, USA}
\altaffiliation{e-mail: leej@ornl.gov}
\author{Istvan K\'ezsm\'aki}
\affiliation{Department of Physics, Budapest University of Technology and Economics and 
MTA-BME Lend\"ulet Magneto-optical Spectroscopy Research Group, 
1111 Budapest, Hungary}
\author{Randy S. Fishman }
\affiliation{Materials Science and Technology Division, Oak Ridge National Laboratory, Oak Ridge, Tennessee, 37831, USA}

\begin{abstract}

Despite its great technological importance, the magnetoelectric (ME) couplings in \BF are barely understood.  
By using a first-principles approach, we uncover the {\it dynamic} ME couplings of the long-range spin-cycloid in BiFeO$_3$. 
Based on a microscopic Hamiltonian, our first-principles approach disentangles the hidden ME couplings 
due to spin-current and exchange-striction. 
Beyond the spin-current polarization governed by the inverse Dzyaloshinskii-Moriya interaction \cite{iDM}, 
various spin-current polarizations derived 
from both ferroelectric and antiferrodistortive distortions 
cooperatively produce the strong non-reciprocal directional dichroism or the asymmetry in the absorption of counter-propagating light in \BP.
Our systematic approach can be generally applied to any multiferroic material, 
laying the foundation for revealing hidden ME couplings on an atomic scale  
and for exploiting optical ME effects in the next generation 
of technological devices such as optical diodes. 

\end{abstract}


\maketitle

The heroic characteristics of \BF, i.e. its room-temperature ferroelectric ($\TC \approx$ 1100 K \cite{teague70}) 
and magnetic ($\TN \approx$ 640 K \cite{sosnowska82}) transitions 
and large ferroelectric polarization \cite{lebeugle07} below $\TC$, 
have unexpectedly hampered our understanding of the magneto-capacitance effects driven by spin ordering below $\TN$. 
Because \BF is a type-$I$ multiferroic, its spin-driven polarizations and magnetoelectric (ME) behavior
are veiled by a large preexisting FE polarization. 
Despite a great deal of effort \cite{kadomtseva04, tokunaga10, park11, sosnowska82, lebeugle08, rama11a, sosnowska11}
and the strong ME effects revealed by recent neutron-scattering \cite{lee13} and Raman-spectroscopy \cite{rov10} measurements, 
little is known about the microscopic origins of the spin-driven polarizations and ME couplings in \BP. 

Due to the lack of spatial inversion and time reversal symmetries in multiferroics, 
the intimate coupling between spins and local electric dipoles can 
give rise to strong ME effects \cite{fiebig05}. 
Such ME effects, mostly studied in the static limit so far, 
can resonantly be enhanced at the so-called ME spin-wave excitations characterized 
by a coupled dynamics of spins and local electric dipoles \cite{fiebig05}. 
Non-reciprocal directional dichroism (NDD) or the difference in the absorption of counter-propagating light beams 
has proven to be a powerful tool to investigate intrinsic ME couplings 
in several multiferroics \cite{kezsmarki11,takahashi12,bordacs12,miy12,szaller13}. 

\BF has two distinctive structural distortions that eliminate inversion centers 
and can couple to the electric component of light. One is the ferroelectric (FE) distortion ($\Gamma_4^-$[111]), 
which breaks global inversion-symmetry (IS), 
and the other is the antiferrodistortive (AFD) octahedral rotation ($R_4^+$[111]), 
which breaks the local IS between nearest neighbor spins. 

Using a first-principles approach based on a microscopic Hamiltonian, we show that all ME couplings are microscopically driven 
by a distinctive combination of these two inherent structural distortions. 
Four spin-current polarizations associated with the FE and AFD distortions 
cooperatively induce the strong NDD in \BP. 
This type of study of dynamical or optical ME effects is especially powerful for 
leaky ferroelectrics where static magneto-capacitance measurements are not feasible and for 
type-$I$ multiferroics such as \BF where 
the evaluation of static magneto-capacitance data is not straightforward due to
the large preexisting FE polarization of roughly $90 \mu$C/cm$^2$ [\onlinecite{lebeugle07}]. 

\vspace{4 mm}

\noindent{\bf 1. Microscopic spin-cycloid model for {\textbf {\emph{R}}3{\textbf {\emph{c}} BiFeO$_3$.}}}
\vspace{4 mm}

The FE and AFD distortions each creates its own Dzyaloshinskii-Moriya (DM) interaction, $\Df$ and $\Da$. 
By including all magnetic anisotropies governed by the FE and AFD distortions, 
the spin Hamiltonian can be written as 
\begin{eqnarray}
&&{\cal H} = \;{\cal H^{\rm SC}_{\rm FE}}\; +\;{\cal H^{\rm SC}_{\rm AFD}}\; +{\cal H^{\rm EX}} + {\cal H^{\rm SIA}} 
\end{eqnarray}
\begin{eqnarray}
&&{\cal H^{\rm SC}_{\rm FE}}= \sum_{\langle i,j\rangle }  \Df \cdot (\vS_i\times\vS_j) \\
&&{\cal H^{\rm SC}_{\rm AFD}}=\sum_{\langle i,j\rangle } \, (-1)^\ni \Da \cdot  (\vS_i\times\vS_j)\\
&& {\cal H^{\rm EX}}=-J_1\sum_{\langle i,j\rangle }\vS_i\cdot \vS_j -J_2\sum_{\langle i,j \rangle'} \vS_i \cdot \vS_j  \\
&&{\cal H^{\rm SIA}}=-K\sum_i (\vS_i \cdot \zp )^2 ,
\label{Ham}
\end{eqnarray}
where $\langle i,j\rangle$ and $\langle i,j\rangle'$ represent 
nearest and next-nearest neighbor spins, respectively.  
The FE polarization lies along $\zp = [1,1,1]$ (all unit vectors are assumed normalized to one). 
Since the FE distortion is uniform, its DM interaction ($\Df$) is translation-invariant. 
By contrast, the translation-odd R$_4^+$[111] AFD octahedral rotation 
requires the coefficient $(-1)^{\ni }$, which alternates from one hexagonal layer $n_i$ to the next, in front of $\Da $.  
The final contribution to the Hamiltonian is the single-ion anisotropy (SIA) proportional to 
the corresponding coefficient $K$. 
SIA favors spin alignment along the FE polarization direction $\zp$.  
Simplified forms for the DM terms $\cal{H}^{\rm SC}_{\rm FE}$ and $\cal{H}^{\rm SC}_{\rm AFD}$ 
are given in Appendix \ref{Heff}.

By ignoring the cycloidal harmonics but including the 
tilt \cite{pyatakov09} $\tau $ produced by $\Da$,
the spin state can be approximated \cite{fishman13b} as
\begin{eqnarray}
\label{syc1}
S_{\xq }(\vR )&=& S (-1)^{n+1} \cos \tau \sin (2\pi \delta r ), \\
\label{syc2}
S_{\yq }(\vR )&=& S \sin \tau \sin (2\pi \delta r ), \\
\label{syc3}
S_{\zq }(\vR )&=&S (-1)^{n+1} \cos (2\pi \delta r ).
\end{eqnarray}
We recall that \cite{fishman13a} $\sin \tau = S_0/S$ where $M_0=2\, \mB S_0$ is the weak FM moment of the AF phase 
along $\yp $ above $H_c$. For moment \cite{tokunaga10,weakFM} $M_0=0.03\,\mB $, $\tau = 0.006$ or 0.34$^\circ $.
By comparison, our result of Local Spin-Density Approximation (LSDA)+$U$ ($U=5$ eV) indicates that $M_0=0.029\, \mB$. 
Because higher harmonics are neglected,
averages taken with the tilted cycloid introduce a very small error of order ${C_3}^2 \approx 2.5\times 10^{-5}$.

\vspace{4 mm}
\noindent{\bf 2. First-principles method}
\vspace{4 mm}

First-principles calculations were performed using density functional theory (DFT) from the VASP code 
within a local spin-density approximation with an additional Hubbard (LSDA+$U$) interaction for the exchange-correlation functional. 
The Hubbard $U$ and the exchange $J_{\rm H}$ were set to $U$ = 5 eV and $J_{\rm H}$ = 0 eV for Fe$^{3+}$,
parameters that were found to be optimal for BiFeO$_3$ \cite{wein12,ed05}. 
We used the projector augmented wave (PAW) potentials \cite{PAW1,PAW2}. 
To integrate over the Brillouin zone, we used a supercell made of a 2$\times$2$\times$2 perovskite units (40 atoms, 8 f.u.), 
3$\times$3$\times$3 Monkhorst-Pack (MP) $k$-points mesh.  To evaluate $\Df$ and $\Da $, we employed a
4$\times$2$\times$2 unit (80 atoms, 16 f.u.) with a 1$\times$3$\times$3 Monkhorst-Pack (MP) mesh. 
The wave functions were expanded with plane waves up to an energy cutoff of 500 eV. 
To calculate exchange interactions ($J_n$), we used four different magnetic configurations 
($G$-AFM, $C$-AFM, $A$-AFM and FM). 
The DM parameters $\Df $ and $\Da $
were estimated by replacing all except 
for four of Fe$^{3+}$ cations with Al$^{3+}$ \cite{wein12} in the 80 atom unit cell.

After obtaining the exchange, DM, aand SIA interactions, 
we calculated their derivatives with respect to 
an applied electric field parallel to a cartesian direction. 
To simulate atomic displacements driven by the applied field ($E_\alpha$) in bulk BiFeO$_3$, 
we calculated the lowest-frequency polar eigenvector from the dynamical matrix
and forcibly move the atoms incrementally 
from the ground state ($R3c$) structure. 
The resulting energy difference between the two structures 
are divided by the induced electric polarization ($P^{\rm ind}_\alpha$). 
The major difference in the polar eigenvectors obtained from the dynamic and the force-constant matrix 
arises from the Fe-O-Fe bond angle. The eigenvector of the dynamic matrix 
decreases the bond-angle while the eigenvector of the force-constant matrix increases that angle (Appendix \ref{vector}). 
These opposing tendencies result in distinct ME behaviors in dynamic and static electric fields. 

In the present study we analyzed the dynamic matrix 
to understand the dynamic ME couplings resulting in NDD. 
\begin{eqnarray}
\label{dielectric}
P^{\rm SD}_\alpha=\frac{\partial H}{\partial E_\alpha}
=\frac{\partial P^{\rm ind}_\alpha}{\partial E_\alpha}\frac{\partial{\cal H}}{\partial P^{\rm ind}_\alpha} 
\approx \epsilon\frac{\partial{\cal H}}{\partial P^{\rm ind}_\alpha} 
\end{eqnarray}
To estimate the dynamic spin-driven polarization ($P^{\rm SD}_\alpha$), 
we calculated $\partial{\cal H}/\partial P^{\rm ind}$ from LSDA+$U$ and 
used the dielectric constant of $\epsilon\approx 90$ 
when the electric field is perpendicular to the rhombohedral axis \cite{lobo07}.

\newpage
\noindent{\bf 3. Spin-current polarizations}
\vspace{4 mm}

The change in the $\vS_i \times \vS_j$ cross product modulates 
the Fe-O-Fe bond angle
and produces the spin-driven polarizations \cite{iDM}.
FE and AFD distortions each generates its own spin-current polarizations
associated with the electric-field derivatives of the DM interactions
$\Df $ and $\Da $, respectively.
They are calculated using the procedure explained in Ref.~\cite{lee15}.

\begin{figure*}
\includegraphics[trim = 0mm 0mm 0mm 2mm, width=16.0cm]{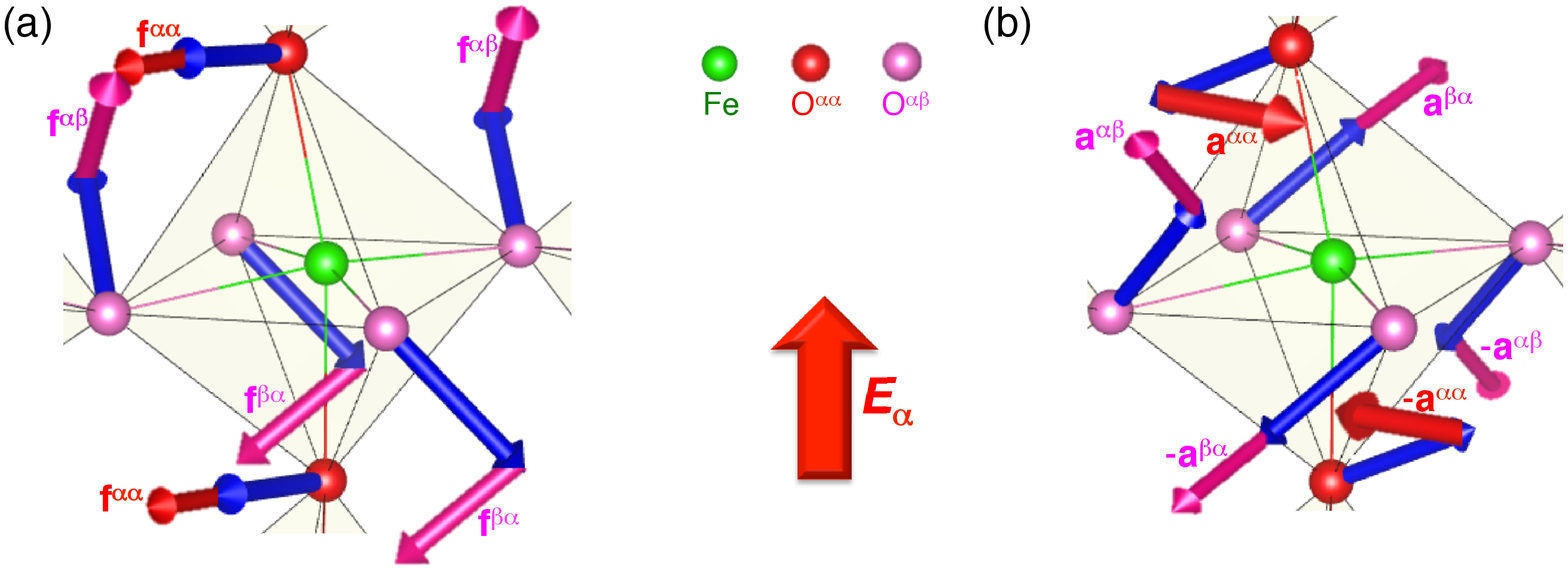}
\caption{Response of Dzyaloshinskii-Moriya (DM) interactions to electric field in $R3c$ BiFeO$_3$.
Blue arrows denote DM vectors without $\vE$ and red arrows denote the change of 
DM with $\vE$. 
  {\bf (a)} FE-induced DM ($\Df$) and its derivative vectors (${\bf f}$) with respect to $\vE $. 
  {\bf (b)} AFD-induced DM ($\Da$) and its derivative vectors (${\bf a}$) with respect to $\vE $. 
The sign of the vectors alternate due to the AFD nature. 
Thick- and light-red arrows denote responses of DM to $\vE $ along the $\alpha $ direction
when spin bonds are parallel (${\bf f}^{\alpha\alpha}$, ${\bf a}^{\alpha\alpha}$) 
and perpendicular (${\bf f}^{\alpha\beta}$, ${\bf a}^{\alpha\beta}$) 
to $\vE $ respectively. 
The size of the arrows is proportional to the magnitudes of the response to $\vE $. 
O$^{\alpha\alpha}$ (O$^{\alpha\beta}$) denotes oxygens along bonds parallel (perpendicular) to $\vE $, respectively.
Bi is not drawn for clarity.}
\label{fna}
\end{figure*}

Hence, the spin-current polarization (SCP) may be written as $\psc = \psc_{\rm FE} + \psc_{\rm AFD}$.
The first SCP is induced by the response of the FE distortion to an external electric field:
\begin{eqnarray}
P^{\rm SC}_{\rm FE,\gamma}=-\frac{\partial {\cal H_{\rm FE}^{\rm SC}}}{\partial E_{\gamma}} 
=
 -\frac{1}{N}\ds\sum_{k, \langle i,j\rangle^{\bf k}} \frac{\partial{\textbf {\emph{D}}^k_{{\rm FE}}}}{\partial E_{\gamma}} 
\cdot \bigl(\vS_i\times \vS_j \bigr) , 
\end{eqnarray}
where $\langle i,j \rangle^{\bf k }$ is a sum over nearest neighbors 
with $\vR_j -\vR_i =a{\bf k} $ and ${\bf k} = \xx $, $\yy $, or $\zz $.  
The electric-field derivatives of the DM interactions 
${\bf f}^{k\gamma}= \partial 
\Df^k /\partial E_{\gamma}$
are given in Appendix \ref{sc} and Tab.~\ref{all}. 
While the derivative (${\bf f}^{\alpha\alpha}$) of $\Df^k$ between spins $\vS_j$ and $\vS_i$ with $\vR_j-\vR_i$ parallel to the electric field
is parallel to $\Df^k$, that (${\bf f}^{\alpha\beta}$) of $\Df^k$ 
between spins with $\vR_j-\vR_j$ perpendicular to the electric field
is perpendicular to $\Df^k$, as shown in Fig.~\ref{fna}. 

In the lab reference frame $\{x,y,z\}$, regrouping terms for domain 2 with $\xp =[1, 0, -1]$ yields 
$P^{{\rm SC}}_{{\rm FE},\alpha} = \sum_{\beta } \Lambda^{\rm FE}_{\alpha \beta } T_{\beta }$ 
with 
\begin{equation}
\underline{\Lambda }^{\rm FE} = 
\left\{\begin{array}{l}
{\bf f}^{xx}\\
{\bf f}^{xy}\\
{\bf f}^{xz}
    \end{array}\right\}
-
\left\{\begin{array}{l}
{\bf f}^{zx}\\
{\bf f}^{zy}\\
{\bf f}^{zz}
    \end{array}\right\}
=
\left(
    \begin{array}{ccc}
    -h & f-g & -f \\
    g & 2h & g \\
    -f & f-g & -h\\
    \end{array} \right),
\end{equation}
where 
\begin{equation}
\label{DDT}
 \vU_1= \frac{1}{N} \sum_{\langle i,j \rangle^{\bf x}} (\vS_i \times \vS_j)
\end{equation} 
and $f={\bf f}_\beta^{\alpha\alpha}$, $g={\bf f}_\beta^{\alpha\beta}$, $h={\bf f}_\gamma^{\alpha\beta}$. 

The second SCP arising from AFD rotations alternates in sign  
due to the alternating AFD rotations along [111]:
\begin{eqnarray}
P^{\rm SC}_{{\rm AFD},\gamma}=-\frac{\partial {\cal H_{\rm AFD}^{\rm SC}}}{\partial E_{\gamma}}
=-\ds \sum_{k, \langle i,j \rangle^{\bf k}}\!\! \frac{(-1)^{n_i}}{N} 
\frac{\partial{\textbf{\emph{D}}^k_{{\rm AFD}}}}{\partial E_{\gamma}} 
\cdot \! \bigl(\vS_i \times \vS_j \!\bigr) . 
\end{eqnarray}
The SCP components ${\bf a}^{k\gamma}=\partial  \Da^k/\partial E_{\gamma}$ 
are evaluated in Tab.~\ref{all}. 
While the derivative (${\bf a}^{\alpha\alpha}$) of $\Da^k$ between spins $\vS_i$ and $\vS_j$
with $\vR_j-\vR_i$ parallel to the electric field is nearly
anti-parallel to $\Da^k$, that (${\bf a}^{\alpha\beta}$) 
of $\Da^k$ between spins with $\vR_j-\vR_i$ perpendicular to the electric field 
is perpendicular to $\Da^k$, as shown in Fig.~\ref{fna}. 

For the spin-cycloid in \BP, the SCP is simplified as (Appendix \ref{simple_sc}), 
\begin{eqnarray}
{\bf P}^{\rm SC}_{\rm AFD}&=& \frac{1}{\sqrt{3}N} \Biggl\{
 \sum_{\langle i,j  \rangle^{\bf x}}(-1)^{n_i}\underline{\Lambda}^{\rm AFD}\cdot (\vS_i \times \vS_{i+x}) 
+ \sum_{\langle i,j \rangle^{\bf y}}(-1)^{n_i}\underline{\Lambda}^{\rm AFD}\cdot (\vS_i \times \vS_{i+y}) \nonumber \\
&&+ \sum_{\langle i,j \rangle^{\bf z}}(-1)^{n_i}\underline{\Lambda}^{\rm AFD}\cdot (\vS_i \times \vS_{i+z}) \Biggr\}
\end{eqnarray}
so that 
\begin{equation}
\underline{\Lambda}^{\rm AFD}={\bf a}^{xx}+{\bf a}^{yy}+{\bf a}^{zz}=
\left(
    \begin{array}{ccc}
    s & t & t \\
    t & s & t \\
    t & t &  s\\
    \end{array} \right),
\end{equation}
where $s=a^{\alpha\alpha}_{\alpha}+2a^{\alpha\beta}_{\beta}$ and 
$t=a^{\alpha\alpha}_{\beta}
+a_{\alpha}^{\alpha\beta}+a_{\gamma}^{\alpha\beta}$ as shown in Tab.~\ref{all}.  

\vspace{4 mm}
\noindent {\bf 4. Two exchange-striction polarizations}
\vspace{4 mm}

The absence of an inversion center between neighboring spin sites also allows the emergence of exchange-striction (ES) polarizations. 
Since the scalar product $\vS_i \cdot \vS_j$ is modified by external perturbations such as temperature, electric or magnetic field, 
the change in the dot product can induces the ES polarizations.
FE and AFD distortions each generates its own ES polarization. 

For symmetric exchange couplings, ES is dominated by the 
response of the nearest-neighbor interaction $J_1$:
\begin{eqnarray}
{\cal H_{\rm ex}}= -\sum_{\langle i,j\rangle }J_1\; \vS_i\cdot \vS_j
=-\sum_{k, \langle i,j\rangle^{\bf k }}J_1^k\; \vS_i\cdot \vS_{j}.
\end{eqnarray}
The two ES polarizations ($P^{{\rm ES}}_{{\rm FE}}$, $P^{{\rm ES}}_{{\rm AFD}}$)
associated with $\vW_1$ and $\vW_2$ are closely related to one another. 
The electric-field derivatives $\underline{\Gamma}$ are given in the cubic coordinate system by 
\begin{equation}
P^{{\rm ES}}_{{\rm FE},\alpha}=-\frac{1}{N}\sum_{\alpha}\frac{\partial{\cal H_{\rm ex}}}{\partial{E_{\alpha}}}
=\sum_{\beta }\Gamma^{\rm FE}_{\alpha \beta }\, W_{1 \beta}    
\end{equation}
\vspace{-1mm}
\begin{equation}
\underline{\Gamma}^{\rm FE} =
\left(
    \begin{array}{ccc}
    C_{\parallel}& C_{\perp} & C_{\perp} \\
    C_{\perp} & C_{\parallel}& C_{\perp} \\
    C_{\perp} & C_{\perp} & C_{\parallel} \\
    \end{array} \right),
  \end{equation}
\vspace{-1mm}
\begin{equation}
 W_{1u}=\frac{1}{N} \sum_{\langle i,j\rangle^{\bf u }} \!\vS_i \cdot \vS_j  ,
\end{equation}
where
$\Cpe=\partial {J_1^{\beta}}/ \partial E_{\alpha}$ ($\beta \neq \alpha$) and $\Cpa= \partial J_1^{\alpha} /\partial E_{\alpha}$   
for spin bonds perpendicular and parallel to the electric field, respectively.  

The AFD octahedral rotation is perpendicular to $\zp$.  Therefore, the ES polarization associated with AFD is also perpendicular to $\zp$ with
\begin{eqnarray}
\label{MS2}
 {\textbf {\emph P}^{\rm ES}_{\rm AFD}}= 
   C_{\rm AFD}\, \zp \times \vW_2 ,
\end{eqnarray}
\begin{equation}
P^{{\rm ES}}_{\rm AFD,\alpha} = \sum_{\beta}\Gamma^{\rm AFD}_{\alpha \beta }\, W_{2\beta } 
\end{equation}
\vspace{-4mm}
\begin{equation}
W_{2u}= \frac{1}{N} \sum _{\langle i,j\rangle^{\bf u }}  (-1)^\ni \,\vS_i \cdot \vS_j  ,
\end{equation}
\vspace{-2mm}
  \begin{equation}
  \underline{\Gamma}^{\rm AFD} =
  \left(
      \begin{array}{ccc}
      0 & -(C_{\parallel}\!-\!C_{\perp}) & C_{\parallel}\!-\!C_{\perp} \\
     C_{\parallel}\!-\!C_{\perp}  & 0 & -(C_{\parallel}\!-\!C_{\perp}) \\
      -(C_{\parallel}\!-\!C_{\perp}) & C_{\parallel}\!-\!C_{\perp} &  0\\
      \end{array} \right).
  \end{equation}
Unlike $W_{1u}$, $W_{2u}$ alternates in sign due to opposite AFD rotations between adjacent hexagonal layers. 

\begin{figure}
\includegraphics[trim = 0mm 2mm 0mm 8mm, width=15.0cm]{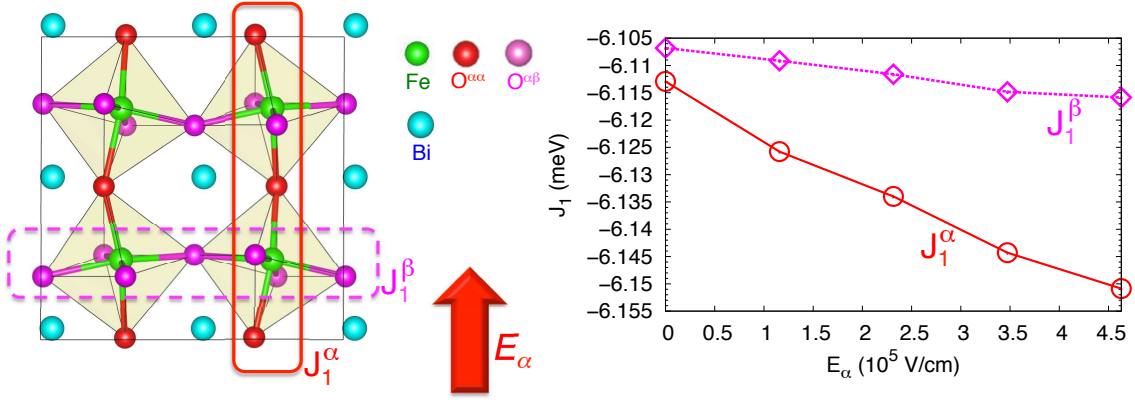}
\caption{ Strong anisotropic response of magnetic exchange ($J_1$) to an electric field.
The slopes of thick and dotted lines represent derivatives of $J_1$ with respect to electric fields
  parallel ($C_\parallel=\partial{J_1^{\alpha}}/\partial{E_{\alpha}}$) 
  and perpendicular ($C_\perp=\partial{J_1^{\beta}}/\partial{E_{\alpha}}$, $\alpha \neq \beta$) 
  to the spin-bond direction calculated from DFT.
}
\label{J1}
\end{figure}

The first ES polarization parallel to $\zp $ with coefficient $C_{\rm FE}=(2\Cpe+\Cpa)$
modulates the FE polarization that already breaks IS above $\TN $. 
The second ES polarization perpendicular to $\zp$ is described by the coefficient $C_{\rm AFD}=\Cpe-\Cpa$. 
The AFD 
distortion affects the bonds between nearest-neighbor spins in the plane normal to $\zp$  
because each oxygen 
moves along $[0, -1,1]$, $[1,0,-1]$, and $[-1,1,0]$, perpendicular to $\zp$. 
Thus, the second ES polarization is associated with atomic displacements perpendicular to $\zp$ and 
parallel to the AFD rotation. 

Figure~\ref{J1} shows a strong anisotropy in the response of magnetic exchange to an electric field. 
$\Cpe$ arises from the change in Fe-O-Fe bond angle due to a polar distortion;
$\Cpa$ arises from bond contraction. As shown in the figure, $\Cpa$ is much more sensitive to an electric field 
than $\Cpe$. 
Since the ME anisotropy $C_{\rm AFD}=\Cpa-\Cpe $ produces an ES polarization associated with AFD, 
the AFD rotation angle is affected by the spin ordering. 
In particular, the negative sign ($C_{\rm AFD}=-250$ nC/cm$^2$) indicates an increase of the rotation angle 
with respect to an increase in the dot product $\vS_i \cdot \vS_j$
because oxygen atoms moving in the AFD plane have a negative effective charge $Z^*_{\rm O}({\rm DFT}) =-3.3e$. 

The anisotropic ES polarization components $\Cpe $ and  $\Cpa$ cooperatively 
induce the ES polarization along $\zp$ under the IS broken by the FE polarization. 
We now obtain a negative $C_{\rm FE}=-350$ nC/cm$^2$ with respect to 
a dynamic electric field 
in contrast to our previous study \cite{lee15} on the response 
to a {\it static} electric field ($C_{\rm FE}=215$ nC/cm$^2$). 
Appendix \ref{vector} shows the different eigenvectors of the dynamic and force-constant matrices. 
Fe moves upward with respect to oxygens in the static regime 
while Fe moves downward in the dynamic regime because  
its mass is much larger than that of oxygen. 
Therefore, a static $\vE $ increases the bond angle 
of Fe-O-Fe (positive $C_{\rm FE}$) but a dynamic $\vE $ 
decreases the bond angle (negative $C_{\rm FE}$) due to 
the Goodenough-Kanamori rules \cite{goodenough} 

\vspace{4mm}
\noindent{\bf 5. Origin of directional dichroism}
\vspace{4mm}

The most stringent test yet for the microscopic model proposed above is its ability 
to predict the NDD, i.e. the
weak asymmetry $\Delta \alpha(\omega)$ 
in the absorption $\alpha(\omega)$ of light when the direction 
of light propagation is reversed.
The absorption of THz light is given by $\alpha (\omega ) = (2\omega /c)\,{\rm{Im}} N(\omega )$
where \cite{miyahara11, miyahara14}
\begin{equation}
N(\omega ) \approx \sqrt{(\underline{\epsilon}_{ii} + \chi^{ee}_{ii} (\omega )) (1+\chi^{mm}_{jj}(\omega ))}\pm  \chi^{me}_{ji}(\omega )
\end{equation}
is the complex refractive index for a linearly polarized beam, 
$\underline{\chi}^{ee}$, $\underline{\chi}^{mm}$ and $\underline{\chi}^{me}$ are the dielectric, magnetic, and 
magnetoelectric susceptibility tensors describing the dynamical response 
of the spin system \cite{kezsmarki11,bordacs12,miyahara11,szaller13} and 
$\underline{\epsilon }$ is the dielectric constant.  
Subscripts $i$ and $j$ refer to the electric and magnetic polarization directions, respectively.
The second term, which depends on the light propagation direction
and produces NDD, is separated from the mean absorption
by writing $N(\omega ) =\bar{N}(\omega )\pm \chi^{me}_{ji} (\omega )$.

Summing over the spin-wave modes $n$ at the cycloidal ordering wavevector $\vQ $,
$\Delta \alpha (\omega ) = (4\omega /c)\, {\rm{Im}} \chi^{me} (\omega ) $ is given by 
\begin{equation}
\label{dal}
\Delta \alpha (\omega ) = \sum_n A_n \, \delta (\omega - \omega_n),
\end{equation}
\vspace{-2mm}
\begin{equation}
A_n= N X \omega_n \, \rm{Re } \Bigl\{ \rho_{n0} \mu_{0n} \Bigr\} ,
\end{equation}
\vspace{-2mm}
\begin{equation} 
\rho_{0n} = \langle 0\vert P^{\rm SD} \cdot \ve /\cV \vert {\it n} \rangle ,\;\;\;\;\;\;\;\;\;\;\;\;
\mu_{0n}= \langle 0\vert \vM \cdot \vh /\mB \vert {\it n} \rangle ,
\end{equation}
where $\vM =(2\mb /N)\sum_i \vS_i$ is the magnetization, $\cV=a^3$ is the volume per Fe site, 
${\bf P}^{\rm SD} /\cV = ({\textbf P}^{\rm ES}+{\textbf P}^{\rm SC})/\cV$
is the net spin-driven polarization
given in units of nC/cm$^2$, and 
\begin{equation}
X= \frac{4\pi \mb }{\hbar } \frac{{\rm nC}}{{\rm cm}^2}= \frac{0.1388}{{\rm cm}}.
\end{equation}
The THz electric and magnetic fields are polarized in the electric ($\ve $) and magnetic ($\vh $) directions, respectively.  

\begin{figure}
\includegraphics[trim = 0mm 0mm 0mm 0mm, width=7.0cm]{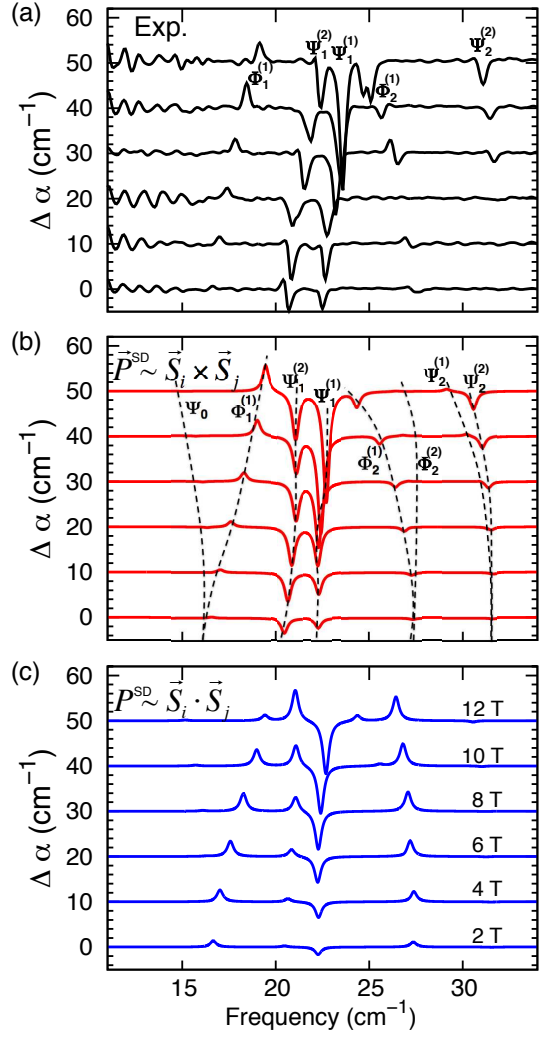}
\caption{
Origin of the strong directional dichroism in BiFeO$_3$.
{\bf (a)} The experimental NDD ($\Delta$ $\alpha$) with static magnetic field from 2 to 12 T
and oscillating electric field along $[1, -1, 0]$. 
The predicted NDD using spin-current {\bf (b)} and exchange-striction {\bf (c)} polarizations.  
$i,j$ denotes nearest neighbors. 
}
\label{NDD}
\end{figure}

\begin{table*}
\caption{\textbf{SD-polarizations from exchange striction, spin current and single-ion anisotropy.}
Shown are the calculated (LSDA+$U$) electric-field derivatives of $J_1$, $\Df, \Da$, and $K$. The upper left and right scripts denote 
the directions of the spin bond and electric field, respectively. 
$f^{\alpha\alpha}_{\beta}=-f^{\alpha\alpha}_{\gamma}$, $f^{\alpha\beta}_{\gamma}=-f^{\beta\alpha}_{\gamma}$, and 
$a^{\alpha\alpha}_{\beta}=a^{\alpha\alpha}_{\gamma}$ by $R3c$ symmetry as in Appendix \ref{sc}. 
$\alpha$, $\beta$, and $\gamma$ are in ascending order so that $\epsilon_{\alpha\beta\gamma}=1$.}
\begin{ruledtabular}
\begin{tabular}{cccccccc}
 & \multicolumn{3}{c}{SCP from $\Df$} & \multicolumn{2}{c}{SCP from $\Da$} 
 & \multicolumn{2}{c}{ES polarization from $J_1$} 
  \\
 & $f^{\alpha\alpha}_{\beta}$ & $f^{\alpha\beta}_{\gamma}$ & $f^{\alpha\beta}_{\beta}$  
 & $a^{\alpha\alpha}_{\alpha}$+2$a^{\alpha\beta}_{\beta}$ 
 & $a_{\beta}^{\alpha\alpha}+a_{\alpha}^{\alpha\beta}+a_{\gamma}^{\alpha\beta}$ 
 & $C_{\rm AFD}$ & $C_{\rm FE}$  \\
 \hline                                              
LSDA+$U$ & 9 & 17 & 14 & 17 & $-19$  &$-250$  & $-350$  \\
Directional dichroism  & 36  & 29 & 29 & 28 & $-7.2$ & - &   -   \\
\end{tabular}
\end{ruledtabular}
\label{all}
\end{table*}

For each field orientation and set of propagation vectors $\ve $ and $\vh $,
the integrated weight of every spectroscopic peak at $\omega_n$ is compared with the measured values,
thereby eliminating estimates of the individual peak widths.  
Experimental results for the NDD with field along $\vm = [1,-1,0]$ are plotted 
in Fig.~\ref{NDD}(a) for $\ve = [1,-1,0]$. 
Fits to the NDD are based on the plotted 2, 4, 6, 8, 10, and 12 T data sets.
For each data set ($\vh $ polarizations per field), we evaluate the integrated weights for the 8 modes \cite{nagel13}
$\Psi_0$, $\Phi_1^{(1)}$, $\Psi_1^{(1,2)}$, $\Phi_2^{(1,2)}$, and $\Psi_2^{(1,2)}$ between roughly 12 and 35 cm$^{-1}$.

From the comparison of Figs.~\ref{NDD}(a) and (b), the NDD for $\vm = [1,-1,0]$
is dominated by the two sets of SC polarizations $\vP^{{\rm SC}}_{\rm FE}$ and $\vP^{{\rm SC}}_{\rm AFD}$ associated
with the DM interactions $\Df$ and $\Da$, respectively. 
Tab.~\ref{all} indicates that the fitting results are not significantly changed by including the ES polarizations.
As shown in Figs.~\ref{NDD}(c) and (d), which minimizes $\chi^2$ with respect to the
experimental measurements \cite{Istvan}, ES polarizations by themselves cannot produce the observed NDD.

Comparing our results to the fits to the NDD, the various components of the spin-current polarizations in BiFeO$_3$ 
are captured by first-principles calculations in Tab.~\ref{all}. 
The optical ME effect responsible the NDD is dominated by the spin-current polarizations
and is not strongly affected by the exchange-striction terms. 
This selective feature originates from the nature of the spin dynamics in BiFeO$_3$. 
Due to the very small single-ion anisotropy on the S = 5/2 Fe$^{3+}$ spins, 
each magnon mode can be described as the pure precession of the Fe$^{3+}$ spins:
the oscillating component $\delta{\bf S}_i^\omega$ of the spin on site $i$ 
is perpendicular to its equilibrium direction ${\bf S}_i^0$. 
Since neighboring spins are close to collinear in the long-range spin cycloid of \BP, 
a dynamic polarization is effectively induced by spin-current terms 
such as ${\bf S}_i^0 \times \delta{\bf S}_{i+1}^\omega$. 
However, the dynamic polarization generated by exchange-striction terms 
${\bf S}_i^0 \cdot \delta{\bf S}_{i+1}^\omega$ is almost zero.
The spin stretching modes observed in strongly 
anisotropic magnets \cite{miyahara11,penc12} does not appear in \BP.

Nevertheless, our DFT calculations underestimate the NDD fitting results in Tab.~\ref{all}. 
We can think of five reasons for this underestimation.
First, a larger dielectric constant ($\epsilon$) could produce better agreement 
between DFT and NDD 
since the spin-driven polarizations are proportional to the dielectric constant that enters Eq.~\ref{dielectric}.
Second, consideration of an electrically-induced polarization ($P^{\rm ind}_\beta$, $\beta\neq\alpha$) 
not parallel to electric field ($P^{\rm ind}_\alpha$) 
could improve the results quantitatively. 
Third, higher-frequency polar modes which were not considered here also can affect NDD. 
Fourth, a smaller Hubbard $U$ 
will increase the spin-driven polarizations and improve the agreement with the experimental fits.
Fifth, magnon modes were observed between $\nu=15$ and 40 cm$^{-1}$
while we calculated the ME couplings in the dynamical limit.  The crossover frequency $\omega_c$ between static and dynamical behavior
lies between 0 and the polar phonon at $\omega=78$ cm$^{-1}$.  If $\omega_c$ lies in the middle of the measured frequencies,
then the polarization parameters may differ from the dynamical couplings evaluated here.

\vspace{4 mm}
\noindent{\bf 6. Discussion}
\vspace{4 mm}

Anchoring first-principles calculations to the right microscopic Hamiltonian is crucial
to understand the ME couplings in complex multiferroic systems.
With two sets of spin-current polarizations derived 
from the two distinct structural distortions, 
\BF is a good example of how our atomistic approach works for complex materials 
beyond the simple inverse DM interaction \cite{iDM} with only one spin-current polarization. 

The advantages (large FE polarization, high $\TC$, and $\TN$) of \BF
have also turned out to be major obstacles  
to understanding the ME couplings that produce the spin-driven polarizations below $\TN$. 
Leakage currents and the preexisting large FE polarization at high temperatures 
have hampered magneto-capacitance measurements 
and hidden the spin-driven polarizations. 
Although recent neutron-scattering measurements \cite{lee13} imply a large ES polarization, most other ME polarizations are unknown. 
However, NDD measurements combined with first-principles calculations based on a microscopic model 
reveal the hidden SC-induced polarizations. 
In particular, this approach allows us to disentangle the delicate spin-current polarizations and the hidden 
ES polarizations associated with AFD rotation that cannot be captured in any other way. 
We envision that intrinsic methods such as NDD will reveal hidden ME couplings in many materials 
and rekindle the investigation of type-$I$ multiferroics.  

\vspace{4 mm}
\noindent{\bf Acknowledgements}
\vspace{4 mm}

We acknowledge discussions 
with H. Kim, E. Bousquet, Nobuo Furukawa, S. Miyahara, J. Musfeldt, U. Nagel, S. Okamoto, S. Bord\'acs 
and T. R\~o\~om. Research sponsored by the U.S. Department of Energy, Office of Basic Energy Sciences, 
Materials Sciences and Engineering Division.
I.K. was supported by the Hungarian Research Fund OTKA K
108918.
We also thank Hee Taek Yi and Sang-Wook Cheong for preparation of the BiFeO$_3$ sample.  

\appendix

\section{Simplified form of Dzyaloshinskii-Moriya (DM) interactions.}\label{Heff}

\subsection{FE-induced Dzyaloshinskii-Moriya (DM) interaction.}

Since the FE vectors $\Df^k$ are given by (0, $D_{\rm FE}$, $-D_{\rm FE}$), 
 ($-D_{\rm FE}$, $D_{\rm FE}$, 0), and ($D_{\rm FE}$, $-D_{\rm FE}$, 0) 
between nearest spins along ${\bf x}$, ${\bf y}$, and ${\bf z}$, respectively, the FE-induced DM interaction can be transformed as:
\begin{eqnarray}
{\cal H}_{\rm FE}^{\rm SC} = 
\sum_{\vR_i , \vR_j=\vR_i + \ve_k}  \Df^k \cdot (\vS_i\times\vS_j) =
D_1 \, \sum_{\vR_i , \vR_j=\vR_i + \ve_k} (\zp \times {\bf e}_k/a) \cdot (\vS_i\times\vS_j),  
\end{eqnarray}
where $D_1= D_{\rm FE} \approx 154$ nC/cm$^2$
is now larger by $\sqrt2$ than in previous work \cite{fishman13a}. 

\subsection{AFD-induced Dzyaloshinskii-Moriya (DM) interaction.}

The AFD interactions $\Da^k$ along ${\bf x}$, ${\bf y}$, and ${\bf z}$ can be written 
\begin{eqnarray}
 {\textbf {\emph{D}}_{{\rm AFD}}^x}=B({\bf y}+{\bf z})+A{\bf x},   \\
 {\textbf {\emph{D}}_{{\rm AFD}}^y}=B({\bf z}+{\bf x})+A{\bf y},   \\
 {\textbf {\emph{D}}_{{\rm AFD}}^z}=B({\bf x}+{\bf y})+A{\bf z}.   \\
 \end{eqnarray}
For the magnetic domain with wavevector along $[1,0,-1]$, 
\begin{eqnarray}
{\cal H}_{\rm AFD}^{\rm SC} &=& 
\sum_{\vR_i , \vR_j=\vR_i +\ve_k } \, (-1)^\ni \Da^k \cdot  (\vS_i\times\vS_j)   \nonumber \\
&=& \sqrt3 \sum'_{\vR_i }\zp \cdot  \Bigl\{ B\;\vS_r \times (\vS_{\vR_i +a{\bf x}}+2\vS_{\vR_i+a{\bf y}}+\vS_{\vR_i+a{\bf z}}) 
+A\;\vS_{\vR_i } \times (\vS_{\vR_i  +a{\bf x}}+\vS_{\vR_i +a{\bf z}})\Bigr\}  \nonumber \\
 &+&\sum'_{\vR_i }\yp \cdot  \Bigl\{(B-A) \; \vS_{\vR_i } \times (\vS_{\vR_i +a{\bf x}}-2\vS_{\vR_i +a{\bf y}}+\vS_{\vR_i +a{\bf z}})\Bigr\} \nonumber \\
&\approx& \sqrt3 \sum'_{\vR_i } \zp \cdot  \Bigl\{ B\;\vS_{\vR_i } \times (\vS_{\vR_i +a{\bf x}}+2\vS_{\vR_i +a{\bf y}}+\vS_{\vR_i +a{\bf z}}) 
+A\;\vS_{\vR_i } \times (\vS_{\vR_i +a{\bf x}}+\vS_{\vR_i +a{\bf z}})\Bigr\},  \nonumber  \\
&\approx& \sqrt3 (4B+2A) \sum'_{\vR_i } \zp \cdot (\vS_{\vR_i } \times \vS_{\vR_i +a{\bf y}}) 
\end{eqnarray}
where the primed sum over $\vR_i $ is restricted to either $n_i$ odd or even hexagonal layers.
Because $\vS_{\vR_i +a{\bf x}}-2\vS_{\vR_i +a{\bf y}}+\vS_{\vR_i +a{\bf z}}$  is of order $\delta^2 \sim 2\times 10^{-5}$, the $\zp$ term dominates. 

Previously, the second DM term was written
\begin{eqnarray}
{\cal H}_{\rm AFD}^{\rm SC} &=& 
D_2 \sum_{\vR_i , \vR_j=\vR_i +\ve_k } \, (-1)^\ni \zp \cdot (\vS_i\times\vS_j)  \nonumber  \\
&=& 2\sqrt3 D_2 \sum'_{\vR_i } \zp \cdot  (\vS_{\vR_i } \times \vS_{\vR_i+a{\bf x}}+\vS_{\vR_i} \times \vS_{\vR_i+a{\bf y}}
    +\vS_{\vR_i} \times \vS_{\vR_i+a{\bf z}}
    ) \nonumber   \\
&\approx& 6\sqrt{3} D_2 \sum'_{\vR_i } \zp \cdot  ( \vS_{\vR_i} \times \vS_{\vR_i+a{\bf y}} )
\end{eqnarray}
Therefore, $D_2 = (A+2B)/3 = 0.064$ meV, which is in
excellent agreement with previous determinations of $D_2$ [\onlinecite{fishman13a}]. 

\section {Eigenvectors of dynamic and force-constant matrix responsible for the different C$_{\rm FE}$.}\label{vector}
\begin{figure}[h]
\includegraphics[trim = 0mm 0mm 0mm 0mm, width=15.5cm]{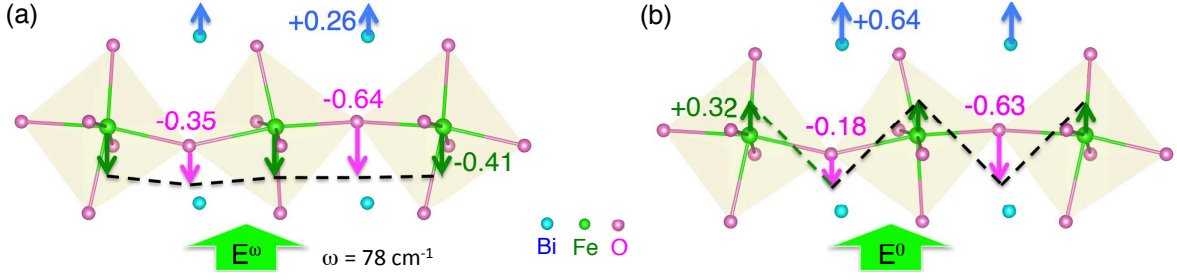}
\caption{Distinct atomic responses to dynamic and static electric fields. 
The lowest-frequency eigenvectors of dynamic matrix (a) and of force-constant matrix (b) are compared. 
Note that the polar displacement in the dynamic limit ($\omega$ = 78 cm$^{-1}$) 
increases the Fe-O-Fe bond angle (dotted line)
while the displacement decreases in the static limit.
} 
\label{AFD}
\end{figure}

We note in the paper that C$_{\rm FE}$ is negative from the eigenmode of the dynamic matrix 
while it is positive from eigenmode of force-constant matrix \cite{lee15}. 
This difference originates from the opposite change of the Fe-O-Fe bond angle. 
The bond angle increases in the static limit (a) while it decreases 
in the dynamic limit (b) ($\omega$ = 78 cm$^{-1}$). 
The different responses to electric field give rise to opposite sign of C$_{\rm FE}$. 

\section{Spin-current polarization components in cubic axis}\label{sc}

Defining ${\bf f}^{k\gamma}=\partial \Df^k/\partial E_{\gamma}$ (f denotes FE distortion), 
\begin{eqnarray}
{\textbf {\emph{D}}_{{\rm FE}}^x}=(0,D,-D),\;\;\;\;\;\; {\textbf {\emph{D}}_{{\rm FE}}^y}=(-D,0,D),\;\;\;\;\;\;  {\textbf {\emph{D}}_{{\rm FE}}^z}=(D,-D,0)   \\
    \vspace{4mm}
{\bf f}^{xx}=(0,f,-f), \;\;\;\;\;\; {\bf f}^{yx} = (-g,0,-h),
 \;\;\;\;\;\;    {\bf f}^{zx} = (g,h,0),           \\
{\bf f}^{xy}=(0,g,h), \;\;\;\;\;\;  
{\bf f}^{yy} =(-f,0,f),
 \;\;\;\;\;\;     {\bf f}^{zy} = (-h,-g,0),          \\
{\bf f}^{xz}=(0,-h,-g),\;\;\;\;\;\;  
{\bf f}^{yz} = (h,0,g),
 \;\;\;\;\;\;     {\bf f}^{zz} = (f,-f,0),         
\end{eqnarray}
where $f \equiv f^{\alpha\alpha}_{\beta}, g\equiv f^{\alpha\beta}_{\beta}$, and $h\equiv f^{\alpha\beta}_{\gamma}$. 

\vspace{4mm}
Defining ${\bf a}^{k\gamma}=\partial \Da^k/\partial E_{\gamma}$ (a denotes AFD distortion), 
\begin{eqnarray}
{\textbf {\emph{D}}_{{\rm AFD}}^x}=(A,B,B), \;\;\;\;\;\; {\textbf {\emph{D}}_{{\rm AFD}}^y}=(B,A,B), \;\;\;\;\;\; {\textbf {\emph{D}}_{{\rm AFD}}^z}=(B,B,A),   \\
    \hspace{16mm}
{\bf a}^{xx}=(a,b,b), \;\;\;\;\;\; {\bf a}^{yx} = (d,c,e),
 \; \;\;\;\;\;    {\bf a}^{zx} = (d,e,c),           \\
{\bf a}^{xy}=(c,d,e),\;\;\;\;\;\;  
{\bf a}^{yy} = (b,a,b),
 \;\;\;\;\;\;     {\bf a}^{zy} = (e,d,c),          \\
{\bf a}^{xz}=(c,e,d), \;\;\;\;\;\; 
{\bf a}^{yz} = (e,c,d),
 \;\;\;\;\;\;     {\bf a}^{zz} = (b,b,a),         
\end{eqnarray}
where $a \equiv a_{\alpha}^{\alpha\alpha}, b\equiv a_{\beta}^{\alpha\alpha}, c\equiv a^{\alpha\beta}_{\alpha}, 
      d\equiv a^{\alpha\beta}_{\beta}$, and $e\equiv a^{\alpha\beta}_{\gamma}$.

\section{Simplification of spin-current polarization (${\bf a}^{\alpha\beta}$) 
from antiferrodistortive DM ($\Da$)}\label{simple_sc}

For domain 2 with $\xp=[1, 0, -1]$, 
\begin{eqnarray}
   \\
{\cal T}_x=\frac{1}{\sqrt3}{\cal T}_{z'}-\frac{1}{\sqrt6}{\cal T}_{y'}  \;\;\;\;
{\cal T}_y=\frac{\sqrt6}{3}{\cal T}_{y'}+\frac{1}{\sqrt3}{\cal T}_{z'}  \;\;\;\;
{\cal T}_z={\cal T}_x  \;\;\;\;\; \Big({\cal T}_k \equiv \frac{3}{N}\sum_i (-1)^{n_i} (\vS_{i} 
    \times \vS_{i+k})\Big)
\end{eqnarray}

The spin-driven polarization associated with $\Da$ is 
\begin{eqnarray}
P^{\rm SC}_{x}&=& {\bf a}^{xx}\cdot{\cal T}_x+{\bf a}^{yx}\cdot{\cal T}_y+{\bf a}^{zx}\cdot{\cal T}_z,   \\
 &=&\frac{1}{\sqrt6}(-{\bf a}^{xx}+2{\bf a}^{yx}-{\bf a}^{zx})\cdot{\cal T}_{y'}
  +\frac{1}{\sqrt3}({\bf a}^{xx}+{\bf a}^{yy}+{\bf a}^{zz})\cdot{\cal T}_{z'}   \\
 &\approx& \frac{1}{\sqrt3}({\bf a}^{xx}+{\bf a}^{yy}+{\bf a}^{zz})\cdot{\cal T}_{z'}
\end{eqnarray}
Similarily, 
\begin{eqnarray}
P^{\rm SC}_{y}&\approx& 
  \frac{1}{\sqrt3}({\bf a}^{xx}+{\bf a}^{yy}+{\bf a}^{zz})\cdot{\cal T}_{z'}    \\
P^{\rm SC}_{z}&\approx& 
  \frac{1}{\sqrt3}({\bf a}^{xx}+{\bf a}^{yy}+{\bf a}^{zz})\cdot{\cal T}_{z'}
\end{eqnarray}
Therefore, in the local frame, 
\begin{eqnarray}  
  P^{\rm SC}_{\xp}&=&P^{\rm SC}_{\yp}=0,  \\
  P^{\rm SC}_{\zp}&=&\frac{1}{\sqrt3}(P^{\rm SC}_x+P^{\rm SC}_y+P^{\rm SC}_z)= 
\frac{1}{3}({\bf a}^{xx}+{\bf a}^{yy}+{\bf a}^{zz})\cdot{\cal T}_{z'}. 
\end{eqnarray}

The polarization matrix used to evaluate the NDD is given by 
\begin{equation}
{\bf a}^{xx}+{\bf a}^{yy}+{\bf a}^{zz}=
  \left(
      \begin{array}{ccc}
      a+2d  & b+c+e & b+c+e \\
      b+c+e & a+2d  & b+c+e \\
      b+c+e & b+c+e & a+2d  \\
      \end{array} \right)
\end{equation}
where
$a+2d = 17$ nC/cm$^2$ and $b+c+e = -19$ nC/cm$^2$ are obtained from first principles as given in Tab.I of the paper.
($a \equiv a_{\alpha}^{\alpha\alpha} = 4.1$ nC/cm$^2$, $b\equiv a_{\beta}^{\alpha\alpha} = -21$ nC/cm$^2$, 
$c\equiv a^{\alpha\beta}_{\alpha} = -6.7$ nC/cm$^2$, 
$d\equiv a^{\alpha\beta}_{\beta} = 6.4$ nC/cm$^2$, and $e\equiv a^{\alpha\beta}_{\beta} = 8.9$ nC/cm$^2$.)

\end{document}